\def\be{\begin{eqnarray}}
\def\en{\end{eqnarray}}

\def\ov{\overline}

\def\CP{{\it CP}~}
\def\pr{{Phys. Rev.}~}
\def\prl{{ Phys. Rev. Lett.}~}

\def\np{{ Nucl. Phys.}~}

\documentclass{ws-ijmpa}

\begin{document}

\markboth{Hai-Yang Cheng} {EFFECTS OF FINAL STATE INTERACTIONS ON
HADRONIC $B$ DECAYS}

%
\catchline{}{}{}{}{}
%

\title{EFFECTS OF FINAL STATE INTERACTIONS ON HADRONIC CHARMLESS $B$
DECAYS\\
}

\author{\footnotesize Hai-Yang Cheng}

\address{Institute of Physics, Academia Sinica\\
Taipei, Taiwan 115, ROC}

\maketitle


\begin{abstract}
Final-state rescattering effects on direct \CP violation in
charmless  hadronic $B$ decays and on the polarization anomaly in
$B\to\phi K^*$ are discussed.

\end{abstract}

\section{Introduction}
Why should we study final-state interactions (FSIs) in charmless
hadronic $B$ decays seriously and urgently  ? It is largely to do
with the recent observation of direct $CP$ violation. A first
confirmed observation of direct $CP$ partial rate asymmetry in
charmless $B$ decays was established recently in $\ov B^0\to
K^-\pi^+$ by BaBar \cite{BaBarKpi} and Belle.\cite{BelleKpi} The
combined BaBar and Belle measurements of $\ov
B^0\to\rho^\pm\pi^\mp$ imply a $3.6\sigma$ direct \CP asymmetry in
the $\rho^+\pi^-$ mode.\cite{HFAG} As for direct \CP violation in
$B^0\to\pi^+\pi^-$, a 5.2$\sigma$ effect was claimed by
Belle,\cite{Bellepipi} but it has not been confirmed by
BaBar.\cite{BaBarpipi}

\begin{table}[h]
\tbl{Comparison of pQCD and QCD factorization (QCDF) predictions
of direct \CP asymmetries (in \%) with experiment. Also shown in
the last column are the FSI modifications to QCDF predictions
taken from [6].}
 {\begin{tabular}{@{}l r r r r @{}}
\toprule Modes & Expt. & pQCD & QCDF & QCDF+FSI  \\
 \colrule
 $\ov B^0\to K^-\pi^+$ & $-11\pm2$ &
 $-17\pm5$  & $4.5^{+9.1}_{-9.9}$ & $-14^{+1}_{-3}$  \\
 $\ov B^0\to\rho^+\pi^-$ & $-48^{+14}_{-15}$ & $-7.1^{+0.1}_{-0.2}$  &
 $0.6^{+11.6}_{-11.8}$ & $-43\pm11$  \\
 $\ov B^0\to\pi^+\pi^-$ & $37\pm24$ &
 $23\pm7$  & $-6.5^{+13.7}_{-13.3}$ & $64^{+3}_{-8}$  \\
 \botrule
\end{tabular}}
\end{table}

Table 1 shows comparison of the model predictions of direct \CP
asymmetries with the world averages of experimental
results.\cite{HFAG} It appears that QCD factorization
predictions\cite{BBNS,BN} for direct \CP violation seem not
consistent with experiment, whereas pQCD results \cite{pQCD} are
in the right ballpark. Recalling that sizable strong phases can be
induced from the annihilation diagram in the pQCD approach by
introducing the parton's transverse momentum, this means that we
have to explore the FSI rescattering phases seriously in QCD
factorization which are unlikely to be small possibly causing
large compound $CP$-violating partial rate asymmetries in
aforementioned charmless decay modes. The sizable \CP asymmetry
observed in $\ov B^0\to K^-\pi^+$ decays is a strong indication
for large direct \CP violation driven by long-distance
rescattering effects.

Besides the above-mentioned \CP violation, there exist some other
hints at large FSI effects in the $B$ physics sector; see
\cite{CCS} for details.

\section{Final State Interactions in charmless $B$ decays}
Based on the Regge approach, Donoghue {\it et al.}\cite{Donoghue}
have reached the interesting conclusion that FSIs do not disappear
in the heavy quark limit and soft FSI phases are dominated by
inelastic scattering, contrary to the common wisdom. A few years
later, it was pointed out by Beneke {\it et al.} \cite{BBNS}
within the framework of QCD factorization that the above
conclusion holds only for individual rescattering amplitudes. When
summing over all possible intermediate states, there exist
systematic cancellations in the heavy quark limit so that the
strong phases must vanish in the limit of $m_b\to\infty$. Hence,
the FSI phase is generally of order ${\cal
O}(\alpha_s,\Lambda_{\rm QCD}/m_b)$. In reality, because the $b$
quark mass is not very large and far from being infinity, the
aforementioned cancellation may not occur or may not be very
effective for the finite $B$ mass. Moreover, the strong phase
arising from power corrections can be in principle very sizable.
Therefore, we will model FSIs as rescattering processes of some
intermediate two-body states with one particle exchange in the
$t$-channel and compute the absorptive part via the optical
theorem.\cite{CCS}

The calculations of hadronic diagrams for FSIs involve many
theoretical uncertainties. Since the one particle exchange in the
$t$ channel is off shell and since final state particles are hard,
form factors or cutoffs must be introduced to the strong vertices
to render the calculation meaningful in perturbation theory. As we
do not have first-principles calculations for form factors, we
shall use the measured rates to fix the unknown cutoff parameters
and then use them to predict direct \CP violation. The results are
shown in the last column of Table 1. We see that direct
$CP$-violating partial rate asymmetries in $K^-\pi^+$,
$\rho^+\pi^-$ and $\pi^+\pi^-$ modes are significantly affected by
final-state rescattering and their signs are different from that
predicted by the short-distance approach.

It is worth stressing a subtle point for $B\to\pi\pi$ decays. The
rescattering charming penguins in $\pi\pi$ are suppressed relative
to that in $K\pi$ modes as the former are Cabibbo suppressed.
Consequently, charming penguins are not adequate to explain the
$\pi\pi$ data: the predicted $\pi^+\pi^-$ ($\sim 9\times 10^{-6}$)
is too large whereas $\pi^0\pi^0$ ($\sim 0.4\times 10^{-6}$) is
too small. This means that a dispersive contribution is needed to
interfere destructively with $\pi^+\pi^-$ so that $\pi^+\pi^-$
will be suppressed while $\pi^0\pi^0$ will get enhanced. This
contribution cannot arise from the charming penguins or otherwise
it will also contribute to $K\pi$ significantly and destroy all
the nice predictions for $K\pi$. In the topological diagrammatic
approach,\cite{Chau} this dispersive term comes from the so-called
vertical $W$-loop diagram ${\cal V}$ in which meson annihilation
such as $D^+ D^-\to\pi\pi$ occurs.


\section{Polarization Anomaly in $B\to\phi K^*$}
For $B\to V_1V_2$ decays with $V$ being a light vector meson, it
is expected that they are dominated by longitudinal polarization
states and respect the scaling law: $1-f_L={\cal O}(m_V^2/m_B^2)$.
However, a low value of the longitudinal fraction $f_L\approx
50\%$ in $\phi K^*$ decays was observed by both BaBar
\cite{BaBarVV} and  Belle \cite{BelleVV}. This polarization
anomaly poses an interesting challenge for any theoretical
interpretation.\cite{Kagan,Colangelo}

Since the scaling law is valid only at short distances, one can
try to circumvent it by considering the long-distance rescattering
contributions from intermediate states $D^{(*)}D_s^{(*)}$. The
large transverse polarization induced from $B\to D^*D_s^*$ will be
propagated to $\phi K^*$ via FSI rescattering. Furthermore,
rescattering from $B\to D^*D_s$ or $B\to DD^*_s$ will contribute
only to the $A_\bot$ amplitude. Recently, we have studied FSI
effects on $B\to VV$. While the longitudinal polarization fraction
can be reduced significantly from short-distance predictions due
to such FSI effects, no sizable perpendicular polarization is
found owing mainly to the large cancellations occurring in the
processes $B\to D_s^* D\to\phi K^*$ and $B\to D_s D^*\to\phi K^*$
and this can be understood as a consequence of \CP and SU(3)
symmetry. Our result is different from a recent similar study in
\cite{Colangelo}. To fully account for the polarization anomaly
(especially the perpendicular polarization) observed in $B\to \phi
K^*$, FSI from other states or other mechanism, e.g. the
penguin-induced annihilation, may have to be invoked.

The same FSI mechanism will also induce sizable transverse
polarization in $B\to\rho K^*$ decays. We found that $f_L(\rho
K^*)$ is reduced to about 60\% which is consistent with data for
$\rho^\pm K^{*0}$ but not for $\rho^0 K^{*\pm}$. This should be
clarified experimentally.

\section{Acknowledgments}

I am grateful to Chun-Khiang Chua and Amarjit Soni for very
fruitful collaboration.

\end{document}